# SOFTWARE ARCHITECTURE FOR FIJI NATIONAL UNIVERSITY CAMPUS INFORMATION SYSTEMS

Bimal Aklesh Kumar

*Department of Computer Science and Information Systems*
*Fiji National University*
*Fiji Islands*
*bimal.kumar@fnu.ac.fj*

**Abstract**

Software Architecture defines the overview of the system which consists of various components and their relationships among the software. Architectural design is very important in the development of large scale software solution and plays a very active role in achieving business goals, quality and reusable solution. It is often difficult to choose the best software architecture for your system from the several candidate types available. In this paper we look at the several architectural types and compare them based on the key requirements of our system, and select the most appropriate architecture for the implementation of campus information systems at Fiji National University. Finally we provide details of proposed architecture and outline future plans for implementation of our system.

## 1. INTRODUCTION

With the advancement of the information and communications technology, variety of software applications have been developed and utilized at many groups, societies and organizations. In educational establishments such as universities campus information systems (CIS) has become extensively used. CIS is a unified system that provides a single point of access to all secure administrative systems at higher education sector. These systems include, but are not limited to, student registration and enrollment, student and employee data, course work and exam information, program information, financial information, human resource information, accommodation, and many more as required by the institute [8].

Prior to the merger and formation of the FNU and due to the autonomous operations of these colleges, at least three different campus information systems existed. The university faced considerable amount of difficulties with these systems. These difficulties included accessibility, scalability, performance, flexibility and integration. In order to address these difficulties, we proposed FNU-CIS to be built on modern code base with modern databases. It should be easily accessible to all the students and the staff of FNU through the local intranet or via World Wide Web. The design should be such where subsequent modification is limited as possible to least cost effect components and should not result in chain reaction of compensating modification hence making it easier to add more functionality in future. It should easily integrate with other systems used in the university such as finance and human resource [8]. These types of software systems are very complex to develop and maintain, it should be developed with the adequate level of performance abilities therefore selection of appropriate software architecture is very important to avoid much of rework, save time and resources.

Software architecture of a computing system is the structure or structures of the system, which comprises software elements, the externally visible properties of those elements and the relationships among them. Like any another complex structure, software must be built on solid foundation. Failing to consider key scenarios, failing to design for common problems, or failing to appreciate the long-term consequences of key decisions can put your application at risk. Modern tools and platforms help you to simplify the task of building applications but they do not replace the need to design your application carefully, based on your specific scenario and requirements. The risks exposed by poor architecture include software that is unstable, is unable to support existing and future business requirements or is difficult to deploy and manage in the production environment [6].

In this paper we look at several architectural styles, compare them based on the key requirements of our system and select the most appropriate architectural style for the implementation of FNU-CIS. Finally we design prototype software architecture for the implementation of proposed FNU-CIS and outline future plans research and implementation of FNU-CIS.

## 2. OVERVIEW OF FNU-CIS

The newly designed system will primarily have three main groups of users; students, academic staff and administration staff. Students include both prospective and current students. Academic staff includes tutors, assistant lecturers, lecturers, senior lecturers and professors. Administration staff includes deans, head of school/department and academic services staff.

*Students*
The students can perform various tasks using the system such as application to study, view/update student profile, program details, apply for graduation, enrollment, timetable, academic history, course work and finance. The current student and prospective students can apply to study at FNU using an online application form. The profile option provides them with choice to view and update their profile such as postal address, residential address, home and mobile phone number. The program details option lists the units that students are required to complete for their program of study. The graduation menu allows the students to apply for graduation online. The enrollment option allows the student to register for courses. The timetable option allows the student view the class and final exam time table. The academic history option lists all the units the student has completed so far. The course work option displays the student's course work for the current semester. The finance option allows the students to view their invoices and make fees payments online through credit cards

*Academic Staff*
The academic staff will be given option such as staff profile, enrolment, course work, class list and HR. Staff profile option allows the staff members to view and update their profile.  The enrollment option allows academic staff members to enroll students for any unit that is activated for particular term of study and the student is not meeting the prerequisite. The student details option allows the staff to access the student profile and academic history. The course work option allows the staff to submit the coursework for the units they are teaching. The class list option allows the staff to access the list of the students that are enrolled for a particular unit. The HR option allows the users to access the HR system used by the institute.

*Administration Staff*
The administration will be given option such as profile, student details, unit activation, applications, graduations, enrollment, and reports. The profile menu allows the administration staff to view and update their profile. The student details option allows the administration staff to access the student profile or academic history of the student. The applications option allows them to view new applications and approve or reject these applications. The graduations option lists the details of those students who have applied to graduate and allows the admin staff to approve or reject this request. The unit activation option allows the HOD's to activate units that would be offered at each campus for a particular term of study. The program update option allows the admin staff to approve or reject the students request to change their program or majors. The reports option allows users to download various types of statistical reports for decision making.

### 3. SOFTWARE ARCHITECTURE

Software architecture is the process of defining a structured solution that meets all of the technical and operational requirements, while optimizing common quality attributes such as performance, security and manageability [8]. It involves a series of decisions based on wide range of factors, each of these decisions can have considerable impact on the quality, performance, maintainability and overall success of the application. Architecture focuses on how the major elements and components with in an application are used by or interact with other major elements and components with in the application. The selection of data structures and algorithms or the implementation details of individual components are design concerns [3].

*3.1. Architectural Styles*
Architectural Style sometimes called an architectural pattern provides an abstract framework for a family of systems. It improves partitioning and promotes design reuse by providing solutions to frequently recurring problem. It is also set of patterns and principles that shape an application [2]. The architecture of a software system is almost never limited to a single architectural style but is often combination of architectural styles that make up the complete system [1]. In general there are four key architectural styles: client-server, component based, n-tier and service oriented architecture.

**Client/Server**
It describes architectural distributed systems that involve a separate client and server connected through network. It segregates the system in two applications where client makes requests to the server. In many cases the server is a database with application logic represented as stored procedures [6]. Some of the variation of client/server style include; Peer to Peer (P2P) applications which allow the clients and servers to swap their roles in order to distribute and synchronized files and information across multiple clients. In some applications

4Bimal Aklesh Kumar / Indian Journal of Computer Science and Engineering (IJCSE)servers host and execute applications and services that thin client accesses through a web browser or specialized thin client applications. The benefits of this style include; that data is only stored on the server thus access and updates to the data are far easier to administer, roles and responsibilities of a computing system are distributed among several servers that are known to each other through network this ensures that a client remains unaware and unaffected by a server repair, upgrade or relocation.

**Component Based**
It decomposes application design into reusable functional or logical components that expose well defined communication interfaces containing methods, events and properties. This provides a higher level of abstraction than object oriented design principles and does not focus on issues such as communication protocols and shared state. Components depend on mechanism with in the platform which provides an environment in which they execute often referred to as component architecture [4]. Common types of component used in application include user interface components such as grid and buttons. The benefit of this architectural style include: ease of deployment, reduced cost, ease of development and reusable.

**N tier**
It segregates the functionality into separate segments. It partitions the concerns of the application into stacked groups. It evolved through component based approach generally using platform specific methods for communication instead of a message based approach. N-tier architecture usually have at least three separate logical parts, each located on as separate physical server and each part is responsible for specific functionality. The nth tier only has to know how to handle request from nth + 1 tier, how to forward request to n – 1th tier and how to handle results of this request. Communication between tiers is typically asynchronous in order to support better scalability. This architectural style provides improved scalability, availability, manageability and resource utilization.

**Service Oriented Architecture**
It refers to application that exposes and consumes functionality as a service using contracts and messages. It enables application functionality to be provided as a set of services and the creation of applications that make use of software services. Services are loosely coupled because they are standard based interfaces that can be invoked, published and discovered. It is focused on providing schema and message based interaction with application through interfaces that are application scoped and not component or object based. The main benefits of this architectural style include domain alignment, abstraction, discoverability, portability and rationalization.

*3.2. Comaprison of software Architecture*

In this section we try comparing the architectural styles given above and try to select the most appropriate style to implement our system. We based the comparison on six key requirements of our system; accessibility, security, performance, flexibility, scalability and maintenance.

- Accessibility - how often users can access the system.
- Security - protecting data from unauthorized use.
- Performance - how fast some aspect of a system performs under a particular workload.
- Flexibility - ability to accommodate change in business requirements with minimum modification.
- Scalability - the load of entity should not grow to an unmanageable size the load should be distributed and shared.
- Maintenance - modification of a software product after delivery to correct faults, to improve performance or other attributes of the application.

ISSN : 0976-5166            Vol. 2 No. 2 Apr-May 2011            257

Table 1: Comparison of Software Architecture

|  | **Client/Server** | **Component Based** | **N-Tier** | **Service Oriented Architecture** |
|---|---|---|---|---|
| *Accessibility* | **Medium** – can be accessed through web browsers and terminals. | **Medium** - can be accessed through web browsers and terminals. | **High** - reach more devices such as (PDA, Mobile) due to its minimum requirements as many devices now have browsers. | **Medium** - can be accessed through web browsers and terminals. |
| *Security* | **High** – data is stored on the server, which offers greater security | **High** – components can implement own security | **High** – separated presentation allows to implement security at various layers | **Poor** - web services typically run on exposed, public facing servers, outside an organization's security perimeter. |
| *Performance* | **Poor** – dependence on central server can negatively impact performance | **Poor** – due to localization of components | **High** – layers over multiple physical tiers improve fault tolerance and performance. | **Poor** – due to localization of services. |
| *Flexibility* | **Poor** – data and business logic is combined on the server | **High**- components are designed to have minimum dependences on other components. | **High** – each tier can be managed easily thus flexibility is high. | **High** – loosely coupled and exposed as independent service on network |
| *Maintenance* | **Medium** – the client remains unaware and unaffected by server repair. | **High** – components can be readily substituted with other similar components. | **High** – each tier is independent of other tiers updates and changes can be easily carried out. | **Medium** – can use well known design principles such as OOP. |
| *Scalability* | **Poor –** high reliance on the central server. | **Poor** – due to localization of components | **High** – tiers are based in deployment of layers scaling out an application is stratight forward. | **Poor** – due to localization of services. |

In the table given above we have compared the four architectural styles based on six key requirements of our system. Given the results of the comparison it can be concluded that n-tier architecture is the best available architectural style for the implementation of our system. It is also very important to decide on how the system would be deployed.

### 3.3. Deployment

In building n-tier application you have to select between *distributed* and *nondistributed* deployment. In non distributed deployment, the layers of application reside on single server excerpt for data storage. This approach has simplicity and minimizes the number of physical layers required. In distributed deployment the layers of application reside as separate physical tiers. Tiered distributing organizes the system infrastructure into physical tiers to provide specific server environments optimized for specific operational requirements. We choose to use nondistributed deployment because distributed deployment has several short comings. All layers will be sharing similar resource, one layer can negatively affect other layers when it is under heavy utilization. A distributed approach allows you to configure servers in such a way that can meet the best performance requirements of each layer, it also allows you to apply more stringent security for e.g. you can add firewall in between web server and application server.

Our system would have n-tier deployment as follows.

Figure 1: FNU-CIS Software Architecture

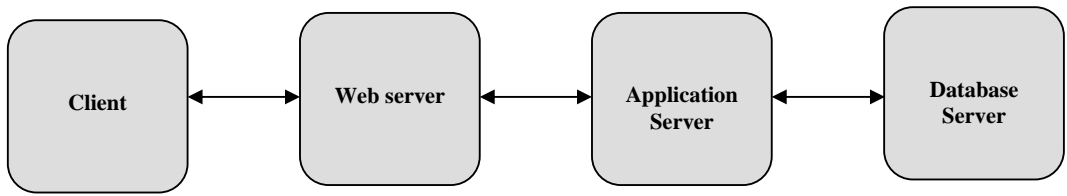

*3.4. Technology*

The leading industry technologies that can be used to implement distributed applications are Microsoft .net and JEE. Microsoft .net is a software framework for Windows operating systems. Java EE is a widely used platform for server programming using Java programming language.

**Microsoft .Net**

It includes a large library, and supports several programming languages which allow language interoperability. The .NET library is available to all the programming languages that .NET supports. The framework's Base Class Library provides user interface, data access, database connectivity, cryptography, web application development, numeric algorithms, and network communications. The class library is used by programmers, who combine it with their own code to produce applications. Programs written for the .NET Framework execute in a software (as contrasted to hardware) environment, known as the Common Language Runtime (CLR). The CLR is an application virtual machine so that programmers need not consider the capabilities of the specific CPU that will execute the program. The CLR also provides other important services such as security, memory management, and exception handling. The class library and the CLR together constitute the .NET Framework. The .NET Framework is intended to be used by most new applications created for the Windows platform.

**Java Platform, Enterprise Edition**

The Java platform (Enterprise Edition) provide functionality to deploy fault-tolerant, distributed, multi-tier Java software, based largely on modular components running on an application server. The platform was known as Java 2 Platform, Enterprise Edition or J2EE until the name was changed to Java EE in version 5..Java EE is defined by its specification. As with other Java Community Process specifications, providers must meet certain conformance requirements in order to declare their products as Java EE compliant. Java EE includes several API specifications, such as JDBC, RMI, e-mail, JMS, web services, XML, etc., and defines how to coordinate them. Java EE also features some specifications unique to Java EE for components. These include Enterprise JavaBeans, Connectors, servlets, portlets (following the Java Portlet specification), Java Server Pages and several web service technologies. This allows developers to create enterprise applications that are portable and scalable, and that integrate with legacy technologies. A Java EE application server can handle transactions, security, scalability, concurrency and management of the components that are deployed to it, in order to enable developers to concentrate more on the business logic of the components rather than on infrastructure and integration tasks.

We chose to implement our system using J2EE and CORBA. J2EE is based on java thus supports cross platform development and availability of world class open source free development environments like Eclipse, and NetBeans lowers the overall development cost. Similarly availability of open source application servers like CORBA along with database servers like MySQL allows both development and deployment to be extremely cost effective as compared to other proprietary application development platforms. CORBA is extremely feature rich supporting numerous programming languages, operating systems and a diverse range of capabilities, such as transactions, messaging and security. Many proprietary middleware technologies are designed with assumptions that developers will build applications using particular middle ware technology so they provide only limited support for integration with other technologies, in contrast CORBA was designed with the goal of making it easy to integrate with other technologies[6]. The flexible server side infrastructure of CORBA makes it feasible to develop servers that can scale from handling up to a number of objects to handling unlimited number of objects [8].

The following technology will be use to develop our system.

Hypertext Markup Language (HTML) - stands for Hypertext Markup Language, is the predominant markup language for web pages. It provides a means to create structured documents by denoting structural semantics for text such as headings, paragraphs, lists, links, quotes, and other items. It allows images and objects to be embedded and can be used to create interactive forms. It is written in the form of HTML elements consisting of "tags" surrounded by angle brackets within the web page content.

Java Server Pages (JSP) - Java Server Pages (JSP) technology enables web developers and designers to rapidly develop and easily maintain, information-rich, dynamic web pages that leverage existing business systems. As part of the Java technology family, JSP technology enables rapid development of web-based applications that are platform independent. JSP technology separates the user interface from content generation, enabling designers to change the overall page layout without altering the underlying dynamic content.

CORBA - CORBA is useful because it enables separate pieces of software written in different languages and running on different computers to work together as a single application or set of services. More specifically, CORBA is a mechanism in software for normalizing the method-call semantics between application objects that reside either in the same address space (application) or remote address space (same host, or remote host on a network).

MySQL - MySQL is a relational database management system (RDBMS) that runs as a server providing multi-user access to a number of databases. Open source projects that require a full-featured database management system often use MySQL. MySQL is also used in many high-profile, large-scale World Wide Web products, including Wikipedia, Google and Facebook.

**4. PROPOSED FNU-CIS SOFTWARE ARCHITECTURE**

N-Tier software architecture for FNU-CIS was designed. It includes client tier, web tier, application server tier and database tier. The client tier (web browser) is implemented using HTML it displays data, collects input from the user and posts it back to the web server for processing. The web tier (web server) it includes server side scripts to serve the request from browser client and generate dynamic content from them. Upon receiving the client request JSP's request from a Java Bean which in turn requests the information from CORBA implemented application servers. Once the Java Beans generate content, JSP's can query and display the content from Java Beans. The application server tier is container for all the components. CORBA is used as the middleware which is implemented using Java language which has CORBA IDL mapping. Database tier is the backend of the system, My SQL is used to implement the database accessed via Java Database connectivity (JDBC).JDBC is an interface that allows java applications to connect to relational databases, when java applications interact with databases JDBC opens database connections and sends SQL commands to query the database [7].

Figure 2: FNU-CIS Software Architecture

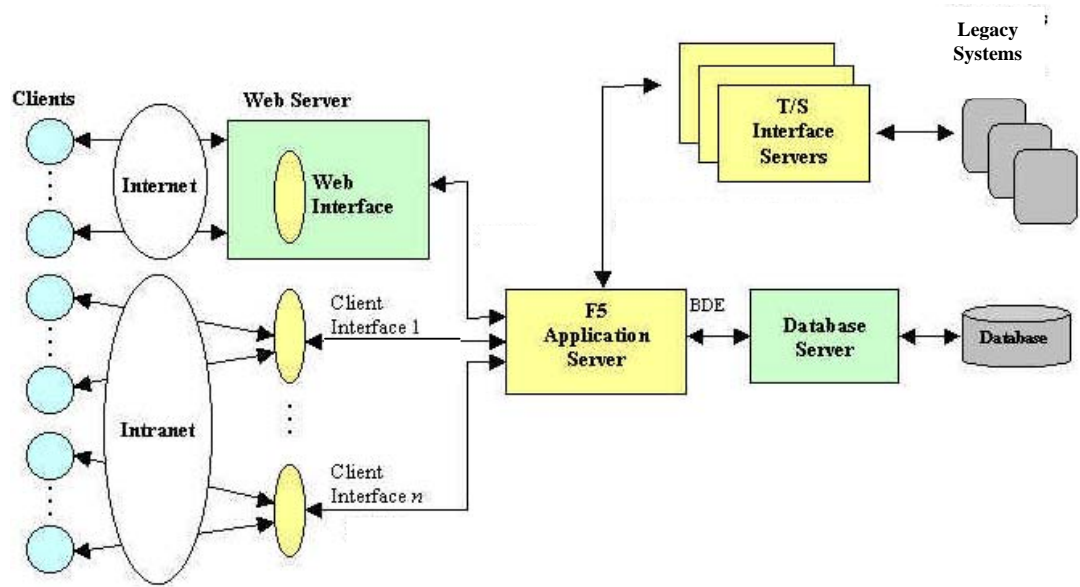

**5.**

**6. ACKNOWLEDGMENT**

I would like to thank Dr. Sharlene Dai, senior lecturer at University of the South Pacific for supervising this research project and Fiji National University for providing financial support to carry out this research.

## 7. CONCLUSION

In this paper we looked at the key architectural styles that can be used to implement our proposed campus information system at Fiji National University. These styles were compared based on the key requirements of FNU-CIS. N-tier architectural style was selected as the most appropriate for implementation of our system. Future work would be carried on exploring further generalization of our software architecture and implementing the software system.